\documentclass[]{aa}

\usepackage{mathptm} 
\usepackage{epsfig} 

\newcommand{\exo}{\mbox{\sl EXOSAT\/}}
\newcommand{\src}{\leavevmode\hbox{4U\,1705--44}}
\newcommand{\sixteen}{\leavevmode\hbox{4U\,1608--52}}
\newcommand{\cyg}{\leavevmode\hbox{Cyg\,X-1}}
\newcommand{\cxt}{\leavevmode\hbox{Cyg\,X-3}}

\begin{document}

\thesaurus{06 (02.01.2; 08.02.1; 08.09.2 \src; 08.09.2 \cyg; 08.09.2
\sixteen; 13.25.5)}

\authorrunning{Berger and van der Klis}
\titlerunning{\src\ and \sixteen\ compared to \cyg}

\title{A comparison of the fast timing behaviour of \src\ to that of \sixteen\ and \cyg} 

\author{M.\ Berger \and M.~van der Klis}

\offprints{M.\ Berger}

\institute{Astronomical Institute ``Anton Pannekoek'', University of 
Amsterdam and Center for High Energy Astrophysics, Kruislaan 403, 1098
SL Amsterdam, The Netherlands}

\date{received <date>, accepted <date>}

\maketitle

\begin{abstract}

We studied the fast timing behaviour of the atoll source \src\ using
the entire \exo\ dataset, four observations covering a total of
230\,000 seconds of 1--20 keV spectral and timing data.  In one of the
observations, \src\ was in a low intensity ``island'' state and had an
unusually hard spectrum. The fast timing analysis of this hard island
state shows a power spectrum very similar to that of black hole
candidates in the `low state', with a flat-topped band-limited noise
component that gradually steepens towards higher frequency.  We
perform for the first time a quantitative comparison of the timing
behaviour of an atoll source in the hard island state (\src) with that
of a black hole candidate in the low state (Cygnus~X-1).  We also
compare the power spectrum of \src\ in the hard island state with
those of the atoll source \sixteen\ in a similar state as reported by
Yoshida et al.~(\cite{yoshida}).  Our results confirm that there are
similarities between the fast timing behaviour of the hard island
states of these atoll sources and the low state of black hole
candidates, yet we also find significant differences in power spectral
parameters; the power spectra of the neutron star systems have a lower
rms amplitude and are less steep.  We find a trend among the neutron
star power spectral properties, in the sense that the lower the
centroid frequency of the fitted Lorentzian is, the higher its
fractional rms amplitude, and the steeper the continuum underneath it.
In our analysis we subtracted the instrumental band-limited noise
component intrinsic to the EXOSAT ME that we found in our previous
work on \cxt, which contaminates the power spectra up to 100 Hz. We
found that this component significantly affects the observed power
spectra. We propose a new method to fit the power spectra of \cyg\ and
other black hole candidates in the low state, that provides a
significantly better fit than previous models.

\keywords{
Accretion, accretion disks 
-- Stars: binaries: close 
-- Stars: individual: \src\
-- Stars: individual: \sixteen\
-- Stars: individual: \cyg\
-- X-Rays: stars
}

\end{abstract}

\section{Introduction}

In 1989, Hasinger and van der Klis (\cite{hk89}) divided the bright
low mass X-ray binaries into two groups, which they called Z sources
and atoll sources after the patterns these sources trace out in the
X-ray colour-colour diagram (CD). The CDs of Z sources show three
branches arranged in a roughly `Z'-shaped pattern, whereas the CDs of
atoll sources show a so-called island and banana, usually
arranged together in one broad, curved branch.  The island state is
characterized by lower count rates, much less motion in the CD and
much stronger band limited noise than the banana state.

\src\ is a burst source that was classified by Hasinger and
van der Klis as an atoll source. From their paper, \src\ appears to be
a special case among the atoll sources in the sense that it has a
`hard island' state (i.e., in its island state it has a harder 1--4.5 keV
spectrum than in its banana state) whereas the other atoll sources
which show both a banana and an island state (4U\,1735--44, 4U\,1636--54
and 4U\,1820--30) have an island state in which the spectrum in the
1--4.5 keV band is softer than in the banana.

The island state of \src\ can be unambiguously identified as such from
its rapid X-ray variability: it shows strong ($\sim 20$ \%) band
limited noise with a cutoff frequency of about 0.3 Hz in this state
(Hasinger and van der Klis, \cite{hk89}, Langmeier et al.,
\cite{paper2}). \src\ is one of the two atoll sources known to exhibit
noise with such strength. The other is \sixteen\ (Yoshida et
al.~\cite{yoshida}).

The motivation for our present analysis is to quantitatively
investigate the suspected similarity (van der Klis, \cite{vdkapjsupp})
both in shape and strength of the noise component found in these two
atoll sources and that in black hole candidates in the low state,
which was up to now based on mostly qualitative considerations. To
this aim, we re-examine the EXOSAT observations of \src, and perform,
for the first time, a systematic, quantitative and direct comparison
with \sixteen\ as well as with the black hole candidate Cyg~X-1.  An
additional reason to carefully look at these data again came from our
earlier work on \cxt\ (Berger and van der Klis, \cite{mezelf}) which
suggested that an instrumental effect of EXOSAT contaminates the ME power
spectra between 0 and 100~Hz. This effect is important for the power
spectra of the banana state especially, and we show that it should be
taken into account.

\section{Observations}

From 1983 until 1986 the instruments of \exo\ were pointed at \src\
four times. A log of the observations is given in Table~\ref{tabobs}.
The lightcurves of the four observations are given in Langmeier et
al.~(\cite{paper1}). The labeling (a) to (d) is identical to the
labeling used by these authors.

\begin{table*}
\caption{Observations of \src\ carried out with \exo\label{tabobs}}
\begin{tabular}{cllllll} \\ \hline
Observation & Start           & End            & OBC    & Time resolution & Source count rate & Nr. of detectors\\
    & (Year/day, Jan 1=1, UT) & (Year/day, UT) & modes  & (1/1024 s)    & (c/s/detector)  & on source \\ \hline
(a) & 1983/211 03:49 & 1983/211 11:06 & HER4, HTR3 & 8          & 203      & 4  \\
(b) & 1985/254 08:50 & 1985/255 13:35 & HER5, HTR3 & 1,8        & 11       & 3,4,7 \\
(c) & 1985/276 17:30 & 1985/277 13:14 & HER5, HTR5 & 1          & 53       & 3,4,7 \\
(d) & 1986/097 07:30 & 1986/098 01:35 & HER5, HTR5 & 1          & 42       & 7 \\
\hline
\end{tabular}
\end{table*}

The \exo\ ME-instrument (Turner et al.\ \cite{turner}, White and
Peacock \cite{pea}) consisted of eight proportional X-ray counters.
In the present analysis, the data from the 1--20 keV argon
proportional counters are used. The data were processed on board the
satellite by the On Board Computer (OBC), which had different modes
emphasizing either spectral (High Energy Resolution or HER) or timing
(High Time Resolution, HTR) information; the OBC modes that were used
and the resulting highest time resolutions are indicated in
Table~\ref{tabobs}.

The array of eight counters of the EXOSAT ME was configured in two
halves of four counters each; one detector failed in 1984.  The half
arrays could both be pointed at the source, or either of them could be
offset by tilting. In this way, background data could be obtained
with one array-half, while making the observation with the other.  In
the rightmost column of Table \ref{tabobs} the number of detectors
pointed at the source in different segments of each observation is
indicated.

\section{Analysis}

\subsection{Timing analysis}

The timing behaviour of \src\ was studied by making fast Fourier
transforms of the high time resolution data, divided into equal length
time intervals. Each time interval was checked for gaps and spikes
(due to telemetry dropouts or other instrumental shortcomings).
Spikes were defined as single-bin excesses over the average count rate
with a probability of chance occurrence from Poisson statistics of
$10^{-8}$ per bin.  Every data segment in which a spike or gap
occurred (only a few percent of the data segments) was excluded from
further analysis.  The data segments affected by X-ray bursts were
also excluded from the analysis.

The resulting Fourier transforms were squared and Leahy normalized
(Leahy et al., \cite{leahy}).  The first step in analyzing these power
spectra was to subtract the Poisson level, a white noise level caused
by counting statistics, which we took to be given by
$P=2(1-\mu\tau_{\rm dead})^2$, with $P$ the Poisson level, $\mu$ the
detected count rate per detector, and $\tau_{\rm dead}$ an
``effective'' instrumental deadtime. In earlier work, we found that
$\tau_{\rm dead}=10.6 \mu$s (Berger and van der Klis, \cite{mezelf}),
and we adopted that value in the present work.  We then renormalized
the Poisson-level subtracted power spectra to (rms/mean)$^2$/Hz
normalization by dividing the Leahy-normalized power spectrum by the
average total observed count rate $I$ appropriate for each spectrum
(e.g.,van der Klis, \cite{kemer}).

We then constructed average power spectra. The power estimates in a
power spectrum of a random process are distributed according to a
chi-squared distribution with 2 degrees of freedom, with a standard
deviation $\sigma$ equal to the power estimate itself, and in practice
the power spectra of X-ray binaries follow this distribution very well
(van der Klis, \cite{cesme}). So, in principle, when averaging power
spectra one should perform a weighted average with weights
$1/\sigma^2$. However, by doing this one would effectively be
assigning each power estimate a weight equal to its inverse value
squared. This would introduce a strong bias in favour of power
estimates that are just statistically low, due to the bin-to-bin
statistical fluctuations. Therefore, it would be preferable to
estimate the weights from the {\it average} power at each frequency
rather than from the power itself, but of course the average power is
not known in advance. An acceptable procedure turns out to be in most
cases to first perform an unweighted average, and use the average
powers resulting from that in setting the weights for the final,
weighted averaging. 

In our case, we have the additional complication that the count rates
$I$ vary strongly between the data sets whose power spectra we are
averaging, causing strong variations in the contribution to the total
power of the Poisson counting noise, which has a value in our
renormalized power spectra of approximately $P_{\rm Poiss} = 2/I$. We
therefore in the first step of the averaging process used weights
$1/\sigma_{\rm Poiss}^2$, where $\sigma_{\rm Poiss} \equiv 2/I$. The
resulting average power spectrum, with powers $\langle P\rangle$, was used to
set the final weights as $1/\sigma^2$, where $\sigma \equiv \langle
P\rangle + 2/I$. These values of $\sigma$ are close to the power
estimates of the final averaged power spectrum, as they should be, and
are much less affected by the bin-to-bin fluctuations in power than if
they would have been estimated from the individual power spectra, thus
avoiding biasing the result in favour of lower power estimates.

The two averaging steps described in the previous paragraph could be
iterated in order to bring the estimates of $\sigma$ even closer to
the final powers, but this proved not to be necessary in our case. The
underlying assumption is, of course, that the instrinsic source
variability within each observation has a constant power spectrum,
which is consistent with what can be derived from inspection of power
spectra of subsets of the data.  To our knowledge this method of
averaging power spectra has not been previously applied in X-ray
timing.

From our analysis of the fast timing behaviour of \cxt\ (Berger and
van der Klis, \cite{mezelf}) it followed that it is likely that an
instrumental effect contaminates the EXOSAT ME argon power spectra
at high frequencies by introducing a band limited noise component with
a cutoff frequency of $\sim$100 Hz and a fractional rms amplitude of
$\sim$3\% of the total (source + background) count rate.

In terms of a (rms/mean)$^2$/Hz normalized power spectrum the
instrumental component was found in the \cxt\ data to be independent
of count rate and given by $P(\nu)=1.44\times 10^{-5} \nu^{5.2\times
10^{-2}} e^{-\nu / 118.2}$, with the frequency $\nu$ in Hz.  The count
rates of \src\ are all in the range of count rates considered by
Berger and van der Klis (\cite{mezelf}), and assuming the same
component is present in the \src\ data, as the \cxt\ analysis
suggests, we subtracted this high frequency noise component from all
power spectra of \src.

\subsection{Spectral analysis}

The spectral behaviour of \src\ was studied by making X-ray
colour-colour diagrams of the available HER data. To do this, the HER
data were divided into four contiguous energy bands with boundaries
0.95, 2.4, 3.4, 5 and 11.6 keV.  We interpolated between the original
ME spectral channels in order to keep these boundaries constant
between observations.  The ratio of the dead-time corrected and
background subtracted count rates in these bands is called an X-ray
colour, the soft colour being the ratio of the 2.4--3.4 and 0.95--2.4
keV count rates and the hard colour the ratio of the 5--11.6 and
3.4--5 keV count rates. We used 256 second integrations per point in
the CD.  The background was measured from slew data or from the offset
array half.

\section{Results}

\begin{figure}
\centerline{\psfig{figure=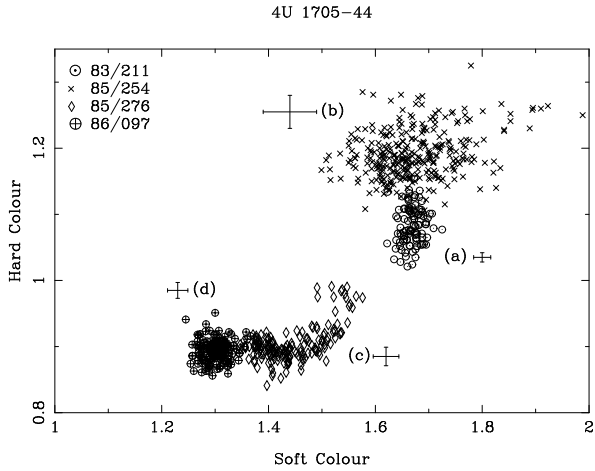,angle=-90,width=8.8cm}}
\caption[ ]{The X-ray colour-colour diagram of \src. Indicated are the
labels as defined in Table \ref{tabobs}, and a typical error bar associated 
with each observation. 
\label{figcd}}
\end{figure}

The colour-colour diagram of \src, included here for reference
purposes only, is shown in Fig.~\ref{figcd}. Different observations
are drawn with different symbols.

\begin{figure}
\centerline{\psfig{figure=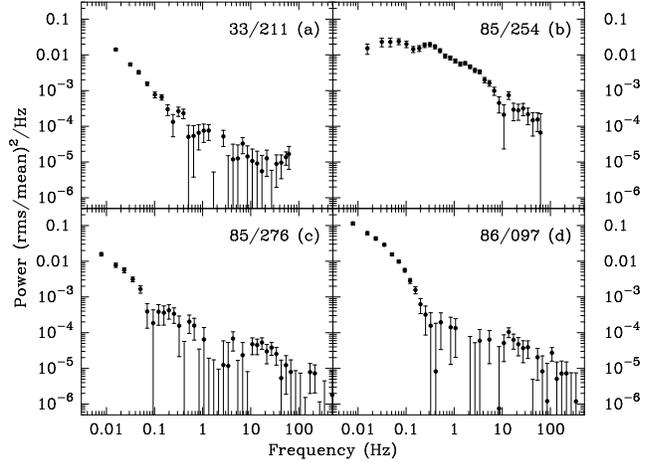,angle=-90,width=8.8cm}}
\caption[ ]{The average power spectra of the four observations of 
\src. The ME instrumental noise component has been 
subtracted in this figure.
\label{figpow}}
\end{figure}

The power spectra of the four observations are drawn in
Fig.~\ref{figpow}.  Observation (c) shows the source in a typical
banana state, with clear very low frequency noise (VLFN) and high
frequency noise (HFN) with an amplitude of only a few percent.
Observation (d) appears very similar, although the VLFN could not be
reliably measured (see below).

Observation (a) shows the source on a very high part of the banana,
with very weak HFN, and VLFN that is stronger than in (c).
Observation (b) is located at the same soft colour, but at slightly
higher hard colour than (a) in the colour-colour diagram. The larger
scatter of the points in the CD of observation (b) is due to the much
lower count rate (see Table \ref{tabobs}). Strong band limited noise
and no VLFN are seen in this observation. In order of ascending count
rate, the observations are (b) $\to$ (d) $\to$ (c) $\to$ (a).  In the
classic `atoll' phenomenology, the island state has a low count rate
and strong variability above 1 Hz. Observation (b) is thus classified
as an island, whereas observation (a) is not.

\subsection{Banana state fits}

The power spectra of the banana state observations (a), (c) and (d)
were fitted with a function consisting of two components: Very Low
Frequency Noise and High Frequency Noise.  The VLFN component was
taken to be a power law $P(\nu) \propto \nu^{-\alpha_1}$.  In
observation (d), made in EXOSAT's last orbit before re-entry into the
Earth's atmosphere, when the count rate was severely influenced by
spacecraft jitter causing variations in collimator transmission, the
VLFN could not be measured.  The HFN component was taken to be an
exponentially cut-off power law $P(\nu)\propto
\nu^{-\alpha_2}\exp({-\nu/\nu_{\rm cutoff}})$.

\begin{table*}
\begin{tabular}{lllllll} \\ 
\multicolumn{7}{c}{Banana state observations} \\ \hline
Obs. & VLFN          & $\alpha_1$      & HFN           & $\alpha_2$     & $\nu_{\rm cutoff}$ & $\chi^2$/dof\\ 
     & \% rms        &                 & \% rms        &                & (Hz)                &  \\ \hline
 (a) & $4.2 \pm 0.2$ & $1.50 \pm 0.05$ & $<1.2$        & --             & $\infty$            & 37.4/30\\
 (c) & $2.7 \pm 0.1$ & $1.32 \pm 0.06$ & $3.4 \pm 0.4$ & $-2.7 \pm 2.0$ & $7 \pm 5$           & 43.0/39\\
 (d) & --            & --              & $3.5 \pm 0.4$ & $-5.4 \pm 3.3$ & $3.2 \pm 2.2$       & 60.2/39\\ \hline
\end{tabular}

\caption[]{
Parameters of the fitted functions to the data where the ME
instrumental noise component has been subtracted. The VLFN rms is
measured from 0.001 to 1 Hz, the HFN rms from 1 to 100. The VLFN of
observation 1986/097 could not be measured due to strong spacecraft
jitter. The upper limit to HFN rms amplitude in (a) is the measured
value plus three times the 1$\sigma$ error obtained from the fit.

\label{tabfitacdwithout}}
\end{table*}

Without subtracting the instrumental component, the measured HFN rms
amplitude was about 5.3\% for observations (c) and (d), and about
3.2\% for observation (a).  The subtraction of the instrumental noise
produces a significant correction to the powers measured in the 1--100
Hz range: The rms amplitudes drop to about 3.4\% for observations (c)
and (d), and for observation (a) now only an upper limit of 1.2\% can
be determined.

\subsection{Island state fits}

\begin{figure}
\centerline{\psfig{figure=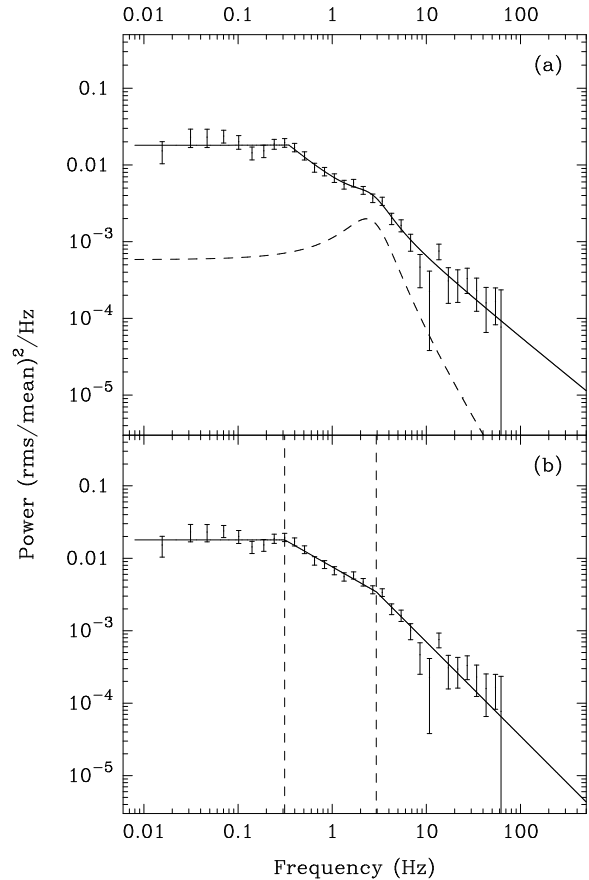,width=8.8cm}}
\caption[ ]{The power spectra of the island state of \src, fitted with
a broken power law plus a Lorentzian near 2 Hz (top panel; the
Lorentzian is indicated by dashed lines) and a three-component broken
power law (bottom panel; the break frequencies are indicated by dashed
lines).\label{figfit}}
\end{figure}

The average power spectrum of the hard island state observation (b)
was fitted with two different functions: A ``once broken power law''
plus a Lorentzian (as used by Yoshida et al.\ in their analysis of
\sixteen), and a ``twice broken power law'' (as used by Belloni and
Hasinger \cite{benh} in their analysis of Cyg~X-1).  The power
spectrum along with the fitted functions is shown in
Fig.~\ref{figfit}, and the parameters of the fitted functions are
shown in Table \ref{tabfit}, where the corresponding values for
\sixteen\ and Cyg~X-1 fits were taken from Yoshida et
al.~(\cite{yoshida}) and Belloni and Hasinger (\cite{benh}).

\begin{figure}
\centerline{\psfig{figure=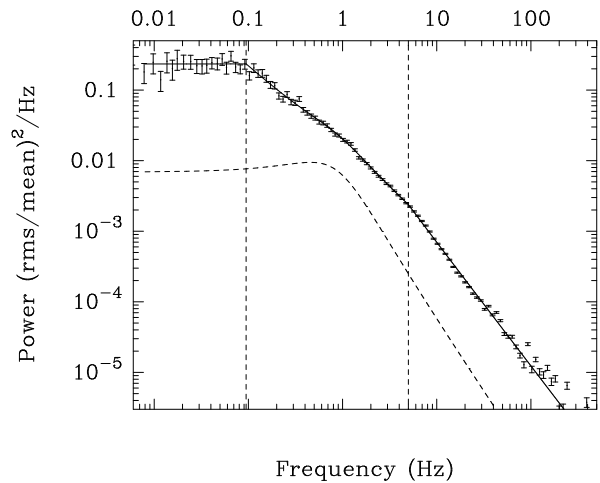,angle=-90,width=8.8cm}}
\caption[ ]{The power spectrum of Cyg~X-1 in its low state, fitted with
a twice broken power law plus a Lorentzian near 0.5 Hz. The break
frequencies are indicated by vertical dashed lines; the Lorentzian
component is indicated separately as well.
\label{figfitcyg}}
\end{figure}

For comparison, a Ginga power spectrum of Cyg~X-1 in its low state
(e.g.~Miyamoto and Kitamoto, \cite{miyanature}) was
fitted with the combination of the two functions: a twice broken power
law plus a Lorentzian located between the two breaks.  Including the
Lorentzian significantly improved the $\chi^2$ of the fit with respect
to the twice broken power law fit, from 450 for 34 d.o.f.~to 352 for
31 d.o.f.  The fit to the low state power spectrum of Cyg~X-1 with its
fit is shown in Fig.~\ref{figfitcyg}.  In Table \ref{tabfitcyg} the
fit results for Cyg~X-1 are given.

We refer to Sect.~\ref{secdiscussion} for a discussion of these results.

\begin{table*}
\begin{tabular}{lll} 
\multicolumn{3}{c}{Once broken power law + Lorentzian} \\ \hline
Parameter & \src\ & \sixteen\ \\ \hline
Broken power law integral (.01-100 Hz)          & $(4.0\pm 0.4) \times 10^{-2}$ & $(3.2 - 9.6) \times 10^{-2}$  \\
Break frequency $ \nu_{\rm break}$(Hz)          & $0.35\pm 0.05$        & 0.2 -- 1.0   \\
Broken power law index, $\nu < \nu_{\rm break}$ & $0.05\pm 0.07$        & 0 (fixed)  \\
Broken power law index, $\nu > \nu_{\rm break}$ & $1.01\pm 0.05$        & 0.8 -- 0.9 \\
Lorentzian integral power                       & $(9.4\pm 3.5) \times 10^{-3}$ & $(2-6) \times 10^{-3}$ \\
Lorentzian centroid frequency (Hz)              & $2.3\pm 0.4$          & 5.1 -- 9.0   \\
Lorentzian FWHM (Hz)                            & $3.0\pm 0.9$          & 4.5 -- 6.1   \\
$\chi^2$/d.o.f.                                 & 20.3/25               & 1--1.6\\ \hline
\end{tabular}

\begin{tabular}{lll} 
\multicolumn{3}{c}{ Twice broken power law} \\ \hline
Parameter & \src\ & Cyg~X-1\\ \hline
Rms integral (\%)                               & $21.3 \pm 0.7$ (0.03-64 Hz) & 28 -- 48 (0.016 -- 4 Hz)\\ 
Low power law index                             & 0 (fixed)             & -0.4 -- +0.2\\
Low break frequency   (Hz)                      & $0.31\pm 0.04$        & 0.05 -- 0.35 \\
Intermediate power law index                    & $0.74\pm 0.06$        & 0.9 -- 1.4\\
High break frequency (Hz)                       & $3.0 \pm 0.6$         & 1.2 -- 5.8\\
High power law index                            & $1.34\pm 0.15$        & 1.4 -- 2.4\\
$\chi^2$/d.o.f.                                 & 27.1/27               & 1.1 -- 3.7 \\
\end{tabular} 
\caption{Left column: Parameters of the fitted once broken power law +
Lorentzian {\em (top)} and the twice broken power law {\em (bottom)} to observation (b) of \src.
Right column: Corresponding parameters for \sixteen\ (Yoshida et al.~1993 ) and Cyg~X-1 (Belloni and Hasinger, 1990).
\label{tabfit}}
\end{table*}

\begin{table}
\begin{tabular}{ll} 
\multicolumn{2}{c}{Cyg~X-1: Twice broken power law + Lorentzian}  \\ \hline
Parameter & Value \\ \hline
Low power law index                             & 0 (fixed)             \\
Low break frequency   (Hz)                      & $(9.5 \pm 0.6) \times 10^{-2} $        \\
Intermediate power law index                    & $1.18 \pm 0.02$        \\
High break frequency (Hz)                       & $5.00 \pm 0.15$         \\
High power law index                            & $1.73 \pm 0.01$        \\
Lorentzian integral power                       & $(2.3 \pm 0.7) \times 10^{-2}$ \\
Lorentzian centroid frequency (Hz)              & $0.42 \pm 0.16$          \\
Lorentzian FWHM (Hz)                            & $1.50 \pm 0.13$          \\
$\chi^2$/d.o.f.                                 & 352/51               \\ 
\end{tabular} 
\caption{Parameters of the fitted twice broken power law + Lorentzian
to the power spectrum of Cyg~X-1 in its low state. 
\label{tabfitcyg}}
\end{table}

We investigated the properties of \src\ in the island state further by
dividing the colour-colour diagram into 3 by 3 rectangular boxes, and
calculating the average power spectrum of all data within each colour
box.  The measured fractional rms variability from 1 to 100 Hz was
consistent with being constant in all 9 boxes (the observed scatter in
the 9 integrated powers was $1.77\times 10^{-4}$, whereas the
predicted scatter from the errors on the individual powers was
$1.71\times 10^{-4}$).

\section{Discussion}
\label{secdiscussion}

The power spectrum of \src\ can be described equally well with a once
broken power law plus a Lorentzian as with a twice broken power law,
and above we have reported results of both fits. On the basis of the
first fit, a direct comparison can be made to \sixteen, as a once
broken power law plus Lorentzian was fitted by Yoshida et al.
(\cite{yoshida}) to the three Ginga observations of the source in its
hard island state. In this quantitative comparison, \src\ turns out to
be indeed quite similar to \sixteen. The break frequency in \src\
(0.35 Hz) is in the range of those found in \sixteen\ (0.2--1.0
Hz). The power law slope of \src\ is a bit steeper (1.0 vs. 0.8--0.9)
and the Lorentzian has a centroid at somewhat lower frequency (2.3
vs.~5--9) than in \sixteen, and a somewhat higher rms amplitude
($\sim9$ vs.~4--7 \%). There is some evidence for a trend in the sense
that the lower the centroid frequency of the Lorentzian is, the higher
its fractional rms amplitude, and the steeper the continuum underneath
it; this holds not only between the two sources, but also between the
three separate observations of \sixteen.

We find that the power spectrum of \cyg\ can be significantly better
described with a twice broken power law plus an additional Lorentzian
than by just a twice broken power law. Comparing our fit of a twice
broken power law plus Lorentzian with the fits of Belloni and Hasinger
(\cite{benh}) to \cyg\ power spectra of just a twice broken power law,
one sees that the two break frequencies we find, at 0.1~Hz and 5~Hz,
are both in the range of the break frequencies they find (0.05--0.35
and 1.2--5.8~Hz). The additional Lorentzian we fit in the power
spectrum of \cyg\ is centered at 0.4 Hz, in between the breaks. This
suggests that the Lorentzian describes additional structure in the
power spectrum, located between the two break frequencies, and does
not strongly affect the basic continuum shape described with the twice
broken power law.  We note that the black hole candidate J0442+32, as
observed with OSSE (Grove et al.~\cite{grove}) shows, in the 35--175
keV band, very clear evidence for the presence of a peak at a similar
frequency (0.23 Hz), superimposed on a similar band limited noise
continuum (see Grove et al.~\cite{grove}, Fig.~2).

To compare \src\ to \cyg, we need make an assumption about the nature
of the correspondence between the power spectral fits to the two
sources. Within the statistics, the data allow the following two
possibilities:

\begin{enumerate}

\item[(i)] The two break frequencies found in the twice broken power
law fit of \src\ (at 0.3 and 3~Hz) correspond to the two break
frequencies found in \cyg, which are at similar frequencies: in our
fit at 0.1 and 5~Hz and in the fits of Belloni and Hasinger in the
ranges 0.05--0.35~Hz and 1.2--5.8~Hz.

\item[(ii)] The Lorentzians and break frequencies found in the once
broken power law plus Lorentzian fits of \src\ and \sixteen\
(centroids 2.3 Hz and between 5 and 9 Hz, breaks 0.35~Hz and
0.1--1~Hz, respectively) correspond to the Lorentzian and lower break
we find in \cyg\ (centroid 0.4~Hz, break 0.1~Hz). Clearly, in this
case the centroid and break frequencies are lower in \cyg\ than in the
two neutron stars; also the strength of the Lorentzian (15\%) is
larger.

\end{enumerate}

In the first case, the implication is that the Lorentzian is missing
from the \src\ fit. We can not exclude that such a component is
present between the two breaks in \src, but it can not have the same
fractional rms amplitude as that seen in \cyg: the 3 $\sigma$ upper
limit is 8.8\%, whereas the amplitude of this component in \cyg\ is
15\%. This means that the ``equivalent width'' (amplitude as fraction
of continuum amplitude) could be approximately the same as in \cyg. In
the second case, the implication is that in \src\ we have not yet seen
the high frequency break. We investigated the possibility of the
presence of such an additional break in the power spectrum of \src\
using the one EXOSAT observation which had a 0.25~ms time resolution,
but the complicated dead time effects of EXOSAT in this mode (see
Tennant \cite{tennant}) and limited statistics prevented us to
constrain the power spectral shape sufficiently well to reach definite
conclusions.

Two conclusions can be drawn irrespective of whether interpretation
(i) or (ii) is true: the neutron star power spectra are less steep,
and correspond to a lower fractional amplitude than those of \cyg. If
(i) the break frequencies of \src\ can be identified with those in
\cyg, then the power law index between the two breaks is 0.7 in \src\
vs. 0.9--1.4 in \cyg, and above the high break frequency 1.3
vs. 1.4--2.4 in \cyg. In case (ii) the power law index above the break
is 0.8--0.9 and 1, respectively, in \sixteen\ and \src, to be compared
with 1.2, steepening further out to 1.7, in \cyg. For \src\ the
fractional rms amplitude of the variability is about 21\% (between
0.01 and 100 Hz), and for \sixteen\ 18--30\% (Yoshida,
\cite{yoshida}). For \cyg\ it is nearly always higher: 28-48\%
(integrated over a smaller frequency interval, 0.016--4~Hz) (Belloni
and Hasinger, \cite{benh}).

Note that in interpretation (ii) \cyg\ extends the trend that was
noted above to exist between \src\ and the three observations of
\sixteen: lower Lorentzian frequency corresponds to higher Lorentzian
amplitude and a steeper continuum.

So, comparing the two atoll (neutron star) sources to the black hole
candidate \cyg\ we find that all these sources have power spectra that
are similar in the sense that they become steeper towards higher
frequency and have fractional rms amplitudes of a few times
0.1. However, the power spectrum of \cyg\ is nearly always
significantly steeper than that in the neutron stars, and the rms
amplitude of the power spectrum is nearly always larger.  Similar
conclusions probably apply to other BHC's in the low state, which have
power spectral characteristics similar to those of \cyg\ (see Miyamoto
et al.~\cite{miya92})

\section{Conclusion}

We studied the similarities in the fast timing behaviour of two atoll
sources in hard island states, and of the black hole candidate \cyg\
in its low state.  In addition to confirming the similarities
discussed previously (van der Klis \cite{vdkapjsupp} and Yoshida
\cite{yoshida}), we performed for the first time a quantitative
comparative analysis of the power spectral features, which showed that
the power spectra of the neutron stars were less steep and had a
smaller rms amplitude than that of \cyg.

Limited statistics prevent us to confirm or reject the presence of an
additional Lorentzian component at lower frequencies (0.3--3.0 Hz) in
the power spectrum of \src, or an additional steepening of the
continuum towards higher frequency, at least one of which would be
predicted on the basis of the analogy to \cyg.  Observations with
future satellites with a larger effective area and faster sampling
speed (like XTE) are required to obtain a more complete picture of the
similarities and differences between the fast timing of black hole
candidates in their low states and atoll sources in hard island
states.

\bigskip
\begin{acknowledgements} 
This work was supported in part by the Netherlands Organization for
Scientific Research (NWO) under grant PGS 78-277.
\end{acknowledgements}

\end{document}